\documentclass{aastex}          
\usepackage{spr-astr-addons}    

\def\arcsec{\hbox{$^{\prime\prime}$}}
\def\SNR{\mbox{{MCSNR~J0536--7038}}}
\newcommand{\SII}{[S\,{\sc ii}]}
\newcommand{\OIII}{[O\,{\sc iii}]}
\newcommand{\Halpha}{H${\alpha}$}
\newcommand{\D}{$^\circ$}
\def\degr{\hbox{$^\circ$}}
\def\arcmin{\hbox{$^\prime$}}

\begin{document}
%
\title{RADIO-CONTINUUM STUDY OF \SNR\ (DEM L249)}

\shorttitle{<Short article title>}
\shortauthors{<Autors et al.>}

\author{L.~M.~Bozzetto\altaffilmark{1}, M.~D.~Filipovi\'c\altaffilmark{1}} 
\email{luke.bozzetto@gmail.com} 

\altaffiltext{1}{University of Western Sydney, Locked Bag 1797, Penrith South DC, NSW 1797, Australia}

\begin{abstract}
We present a detailed radio-continuum study on Australia Telescope Compact Array (ATCA) observations of Large Magellanic Cloud (LMC) supernova remnant (SNR), \SNR. This Type~Ia SNR follows a horseshoe morphology, with a size 32~pc $\times$ 32~pc (1-pc uncertainty in each direction). It exhibits a radio spectrum $\alpha=-0.52\pm0.07$ between $\lambda = 73$ and 6~cm. We report detections of regions showing moderately high fractional polarisation at 6~cm, with a peak value of 71$\pm$25\% and a mean fractional polarisation of 35$\pm$8\%. We also estimate an average rotation measure across the remnant of --237~rad~m$^{-2}$. The intrinsic magnetic field appears to be uniformly distributed, extending in the direction of the two brightened limbs of the remnant.\\
\end{abstract}


\section{Introduction}

Supernova and their remnants play a crucial role in the evolution and structure of the interstellar medium (ISM), as they are responsible for the creation and distribution of the heavier elements and the chemical enrichment of galaxies. The Large Magellanic Cloud (LMC) is a favourable environment for observing supernova remnants (SNRs), due to the relatively close proximity in which it is located (and in turn larger angular size), making it possible to resolve these objects (as 1~pc = ~4\arcsec). One LMC SNR of particular interest is \SNR, a Type~Ia SNR that has prominent Fe emission in its core and is thought to be the result of a `prompt' Type~Ia event \citep{bork06, maggi13}.

\SNR\ was originally observed by \citet{davies76} using the 48-in SRC Schimdt camera, who gave this object the association DEM~L249. They estimate an optical diameter of 180\arcsec$\times$120\arcsec\ making note that this object was very faint and had diffused emission. \citet{mathewson83} find an optical size of 162\arcsec$\times$129\arcsec\ and an X-ray size of $\sim$120\arcsec, an integrated flux density at 408~MHz of 130~mJy and a radio spectral index of $\alpha=-0.53$. \citet{mills84} recorded an integrated flux density measurement at 843~MHz of 850~mJy, updating the spectral index to $\alpha=-0.52$. They also note a horseshoe morphology for which they state that approximately one third of SNRs greater than 30\arcsec\ are. \citet{fusco84} find an ambient density of 0.05~cm$^{-3}$, a shock temperature of 1.5~keV, an age of 6.5$\times$10$^3$ years, total swept up mass of 55$M_{\odot}$ and a shock velocity of 1100~km~s$^{-1}$. No pulsar was found at this position by \citet{manchester85} in their search for short period pulsars. \citet{berk86} measure a radio surface brightness of 20.5~W~Hz$^{-1}$ m$^{-2}$ sr$^{-1}$ using the radio diameter of 30~pc and spectral index of $\alpha=-0.52$. Also measured was a X-ray surface brightness of 32.4~erg~s$^{-1}$ pc$^{-2}$ using the X-ray extent of 39~pc. \citet{chu88} listed the OB association of 560~pc to LH66,69 and listed the source as population type~II? (with `?' indicating uncertain classification). They also made note that there was no detectable CO in the vicinity. \citet{filipovic95} used the Parkes radio telescope and measured integrated flux densities of 25~mJy (4750~MHz) and 34~mJy (4850~MHz). \citet{williams99} listed an average brightness of 0.0010~cts~s$^{-1}$~arc~min$^{-2}$, an X-ray size of 5.2\arcmin$\times$3.3\arcmin\ and stated that the category of this SNR was unclassified as it did not show an unambiguous detection in the ROSAT observations. \citet{hp99} recorded an extent of 29.7\arcsec\ and gave this SNR the association HP[99]~1173. \citet{urosevic05} included \SNR\ in their study of the $\Sigma$-{\it D} relationship. Using a diameter of 39~pc, an integrated flux density of 62~mJy at 1400~MHz (which was extrapolated from the values listed for $\alpha$ and S$_{36}$ by \cite{filipovic98}) and a spectrum of $\alpha=-0.61$, they calculated the surface brightness of this SNR to be 2.0$\times$10$^{-21}$~W/M$^2$~Hz~sr. \citet{blair06} record this SNR as a large faint optical shell SNR which is brighter on the eastern side of the shell. They make note that it may be a mixed morphology class, and that C$_{III}$ is present in its spectra which may indicate that the SNR is on the near side of the LMC. \citet{seok08} estimated an integrated flux of 9~mJy (4800~MHz). An optical extent of 3.1\arcmin$\times$2.3\arcmin\ was recorded by using the MCSNR Atlas as published by Williams (http://www.astro.uiuc.edu/projects/atlas/). \citet{payne08} spectroscopically confirmed nature of this SNR based on the \SII/\Halpha\ ratio. \citet{desai10} recorded an extent of 3.0\arcmin$\times$2.0\arcmin\ and no detection of a YSO for this object nor association with the molecular clouds.

Here, we report on new radio-continuum observations of \SNR. The observations, data reduction and imaging techniques are described in Section~2. The astrophysical interpretation of newly obtained moderate-resolution total intensity and polarimetric images in combination with the existing Magellanic Cloud Emission Line Survey (MCELS) images are discussed in Section~3.

\section{Observations and reduction}

We observed \SNR\ along with various other LMC SNRs via the ATCA on the 15$^\mathrm{th}$ and 16$^\mathrm{th}$ of November 2011, using the new Compact Array Broadband Backend (CABB) at array configuration EW367 and at wavelengths of 3 and 6~cm ($\nu$=9000 and 5500~MHz). Baselines formed with the $6^\mathrm{th}$ ATCA antenna were excluded, as the other five antennas were arranged in a compact configuration. The observations were carried out in the so called ``snap-shot'' mode, totaling $\sim$50 minutes of integration over a 14 hour period. PKS~B1934-638 was used for flux density calibration\footnote{Flux densities were assumed to be 5.098 Jy at 6~cm and 2.736 at 3~cm.}  and PKS~B0530-727 was used for secondary (phase) calibration. The phase calibrator was observed twice every hour for a total 78 minutes over the whole observing session. The \textsc{miriad}\footnote{http://www.atnf.csiro.au/computing/software/miriad/}  \citep{sault95} and \textsc{karma}\footnote{http://www.atnf.csiro.au/computing/software/karma/} \citep{gooch95} software packages were used for reduction and analysis. More information on the observing procedure and other sources observed in this project can be found in \citet{bozzetto2012a, bozzetto2012b, bozzetto2012c, bozzetto2012d,bozzetto2013} and \citet{dehorta2012}.

The CABB 2~GHz bandwidth is a 16 times improvement from the previous 128~MHz, and with the new higher data sampling has increased the sensitivity of the ATCA by a factor of 4. The 2~GHz bandwidth not only aids in high sensitivity observations, but also allows data to be split into channels which can then be used for measuring Faraday rotation across the entire bandwidth, at frequencies close enough that the {\it n}~$\times$~180\D ambiguities prevalent when making an estimate between distant frequencies, are no longer an issue.

\begin{figure}[h!]
\begin{center}
\includegraphics[scale=.35, angle=-90, trim=0 40 0 30,clip]{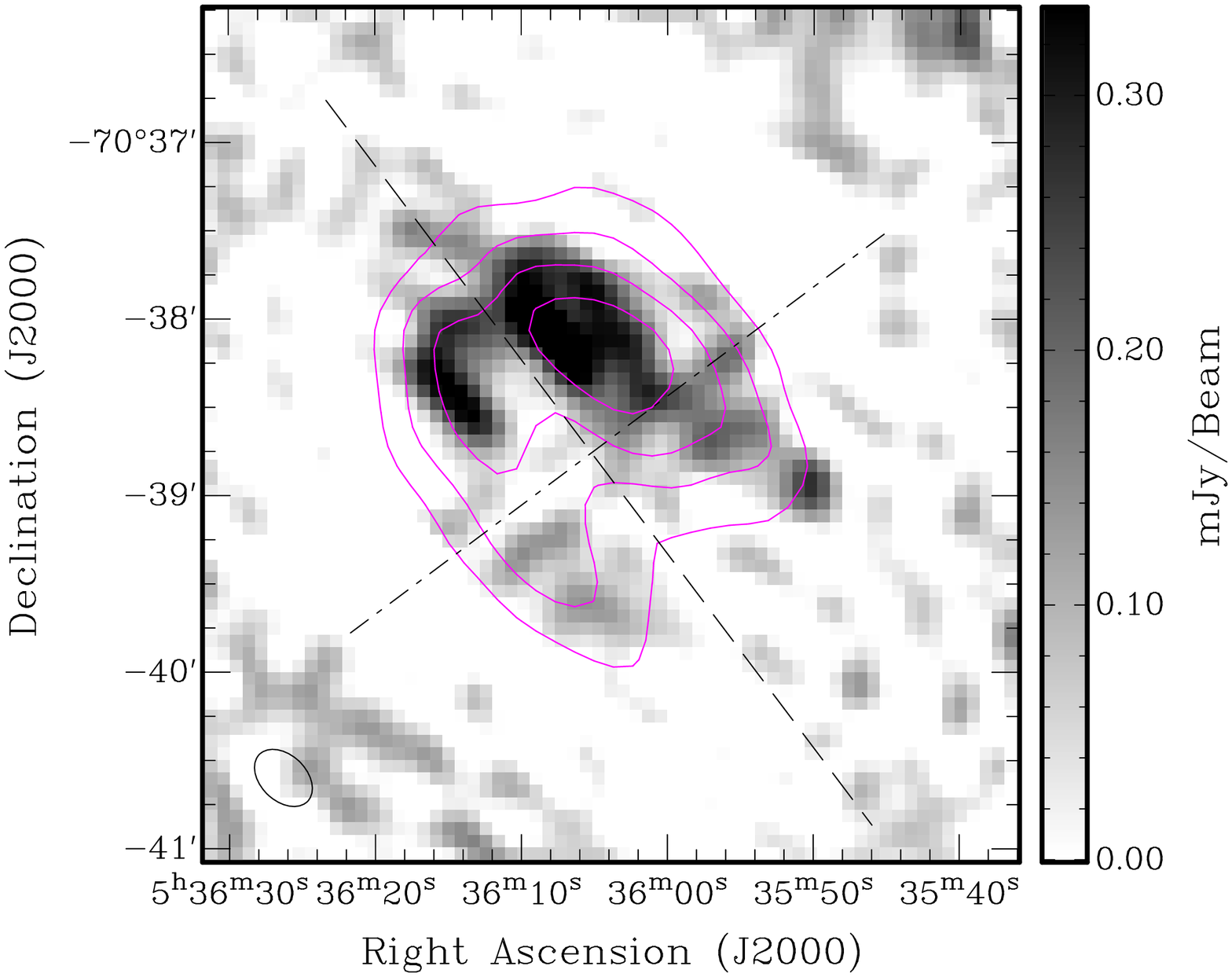}
\includegraphics[scale=.25, angle=-90, trim=0 55 -10 0]{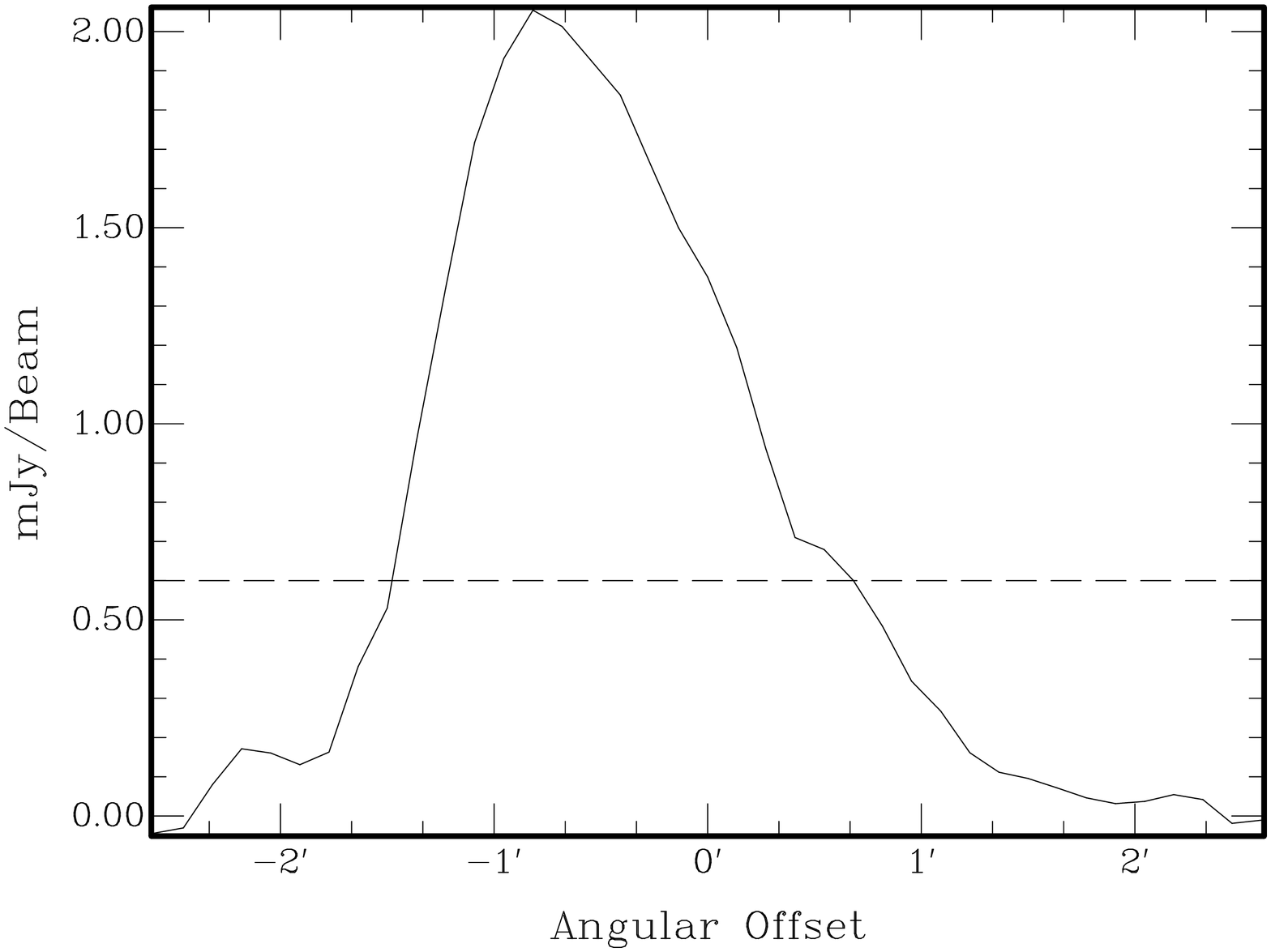}
\includegraphics[scale=.25, angle=-90, trim=0 55 0 0]{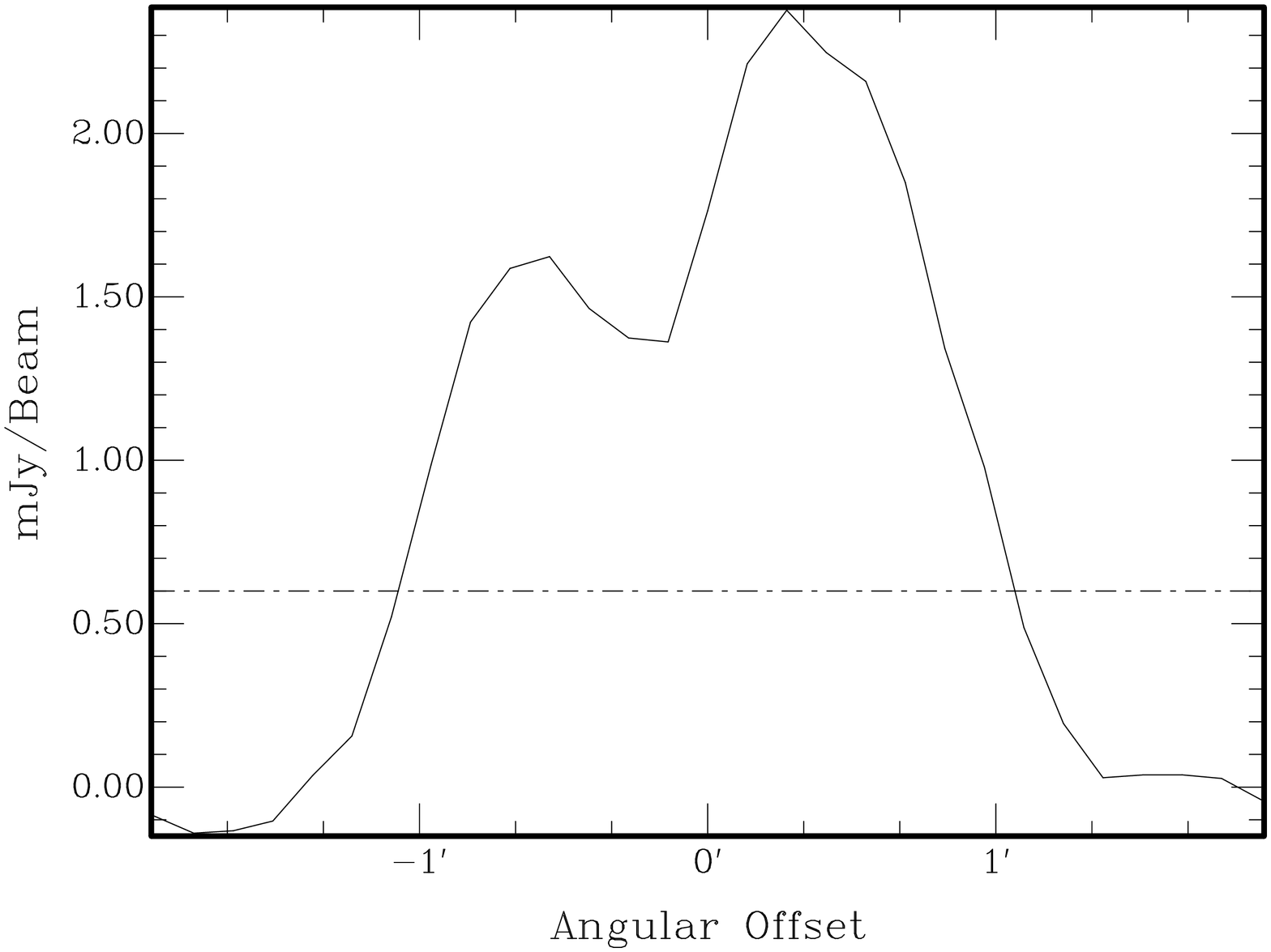}
\caption{The top image shows the 3~cm intensity image of \SNR\  overlaid with 6~cm contours. The contours are 3, 6, 9 \& 12$\sigma$ ($\sigma$ = 0.2~mJy/beam). The ellipse in the lower left corner represents the synthesised beamwidth at 3~cm of 22.4\arcsec$\times$15.7\arcsec. Also superimposed, are the major (NE--SW) and minor (SE--NW) axis. The middle and lower images show the intensity profile of the major and minor axis respectively, with an overlaid line at 3$\sigma$}\label{extent}
\end{center}
\end{figure}

Images were formed using \textsc{miriad} multi-frequency synthesis \citep{sault94} and natural weighting. They were deconvolved with primary beam correction applied. The same procedure was used for both \textit{U} and \textit{Q} stokes parameter maps. 

The 3~cm image (Fig.~\ref{extent}) has a resolution (full width half maximum (FWHM)) of 22.4\arcsec$\times$15.7\arcsec\ (PA=44.5\D) and an r.m.s noise of 0.1~mJy/beam. Similarly, we made an image of \SNR\ at 6~cm (seen as contours in Fig.~\ref{extent}) which has a FWHM of 38.5\arcsec$\times$24.2\arcsec\ (PA=48.1\D) and an estimated r.m.s. noise of 0.2~mJy/beam.

\begin{figure}[h]
\begin{center}
\includegraphics[scale=.45, angle=-90, trim=0 40 0 0]{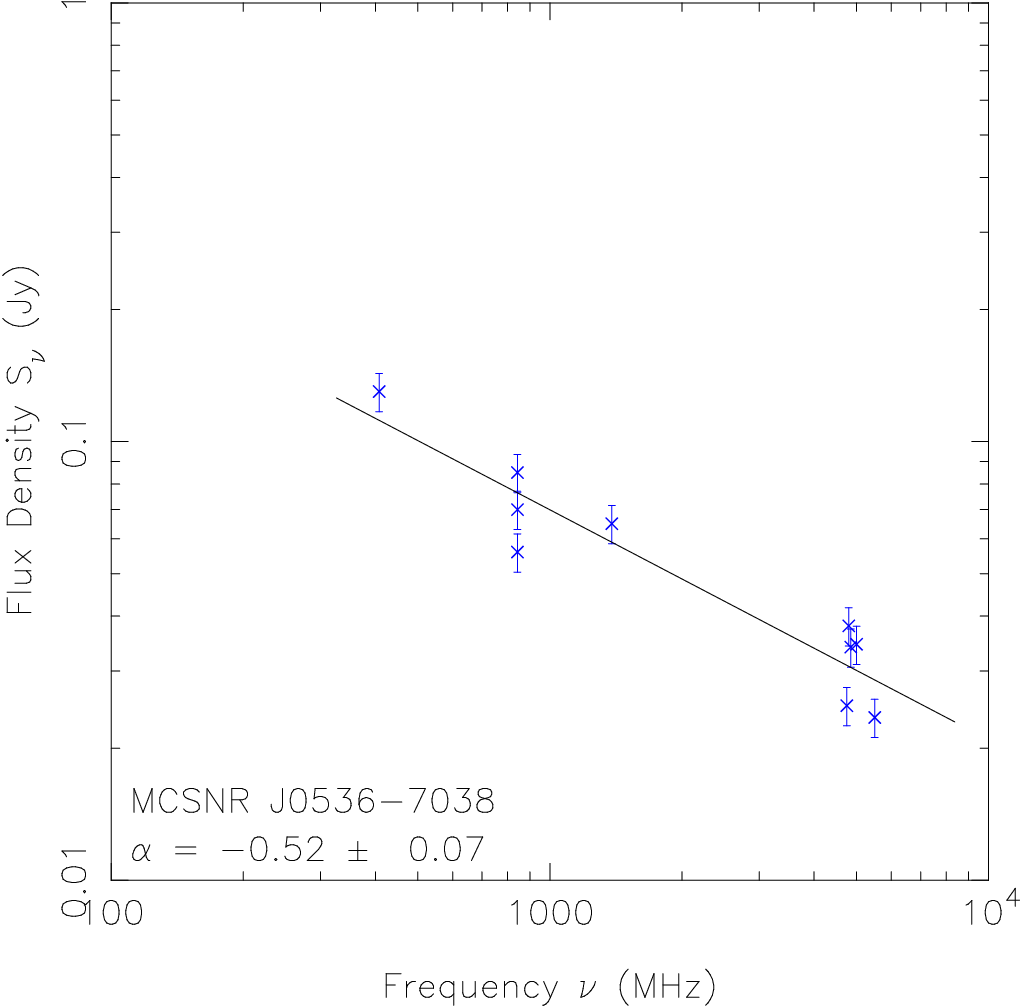}
 \caption{Radio spectrum of \SNR}
 \label{spcidx}
\end{center}
\end{figure}

In addition to using our own observations, we measure the integrated flux density at several frequencies between $\lambda$ = 36 cm and 6~cm, and record these measurements along with previous measurements of this SNR in Table. \ref{tbl-flux}. The 36~cm (843~MHz) measurements come from the Molonglo Synthesis Telescope (MOST; as described in \citealt{mills84}) and Sydney University Molonglo Sky Survey (SUMMS; \citealt{mauch08}) mosaic images. These 36~cm surveys are from two different sets of observations and we point that the MOST data comes with somewhat better {\it uv} coverage and sensitivity. This would be the most likely reason for a difference in flux density estimates (14~mJy or $\sim$20\%) at this wavelength. The 20~cm (1384~MHz) measurement is from the mosaic image described in \citet{hughes07}, while the 6~cm (4800~MHz) measurement comes from a mosaic image published by \citet{dickel10}. Errors in these measurements predominately arose from defining the edge of the remnant. However, we estimate that the error in these measurements are in the order of 10\%. Using these values from Table \ref{tbl-flux}, we estimate a spectral index for \SNR\ of $\alpha=-0.52\pm0.07$ (Fig.~\ref{spcidx}). We note the gap in flux density between our 6~cm (5500~MHz) measurement and the measurement from the 6~cm (4800~MHz) \citet{dickel10} mosaic. This can be explained by the missing short (zero) spacing measurement in interferometery, which is responsible for the large scale, extended emission. The shortest gap in our baseline array of 46~m affected the amount of flux observed, whereas the \citet{dickel10} mosaic incorporated a single dish (and therefore, zero-spacing) measurement from the Parkes radio telescope in addition to ATCA observations. Flux density measurements taken from our 3~cm image were omitted from the spectral index calculation as the effects from short spacing were far more detrimental than at 6~cm, as seen in Fig.~\ref{extent}. Measurements taken from the 3~cm \citet{dickel10} mosaic were also omitted as there was no reliable emission at the location of the remnant.

\begin{table*}
\begin{center}
\caption{Integrated flux density measurements for \SNR.}\label{tbl-flux}
\begin{tabular}{ccccccl}
\hline
$\nu$ & $\lambda$ & r.m.s.      & Beam Size  & S$_\mathrm{Total}$  & $\Delta$S$_\mathrm{Total}$ & Reference \\
(MHz) & (cm)      & (mJy/beam)  & (\arcsec) & (mJy) & (mJy) &\\
\hline 
408     & 73 & 40  & 157.2$\times$171.6 & 130 & 13 & \citet{mathewson83}\\
843     & 36 & --- & 46.4$\times$43.0   & 85  & 9  & \citet{mills84}\\
843$^a$ & 36 & 0.8 & 46.4$\times$43.0   & 70  & 7  & This work\\
843$^b$ & 36 & 1.0 & 47.3$\times$45.0   & 56  & 6  & This work\\
1384    & 20 & 0.6 & 40                 & 65  & 7  & This work\\
4750    & 6  &  8  & 288                & 25  & 3  & \citet{filipovic95}\\
4800    & 6  & --- & 35.0$\times$35.0   & 38  & 4  & This work\\
4850    & 6  & 5   & 294                & 34  & 3  & \citet{filipovic95}\\
5000    & 6  & --- & 258                & 34.5& 3  & \citet{mathewson83}$^c$\\
5500    & 6  & 0.2 & 38.5$\times$24.2   & 24  & 2  & This work\\
\hline
\end{tabular}
\medskip\\
\end{center}
$^a$Uses the MOST mosaic image.\\
$^b$Uses the SUMMS mosaic image.\\
$^c$This value was derived from the 408~MHz flux value from \citet{mathewson83} and their spectral index.\\
\end{table*}

\begin{figure}[h!]
\begin{center}
\includegraphics[scale=.53,trim=14 50 80 50, clip]{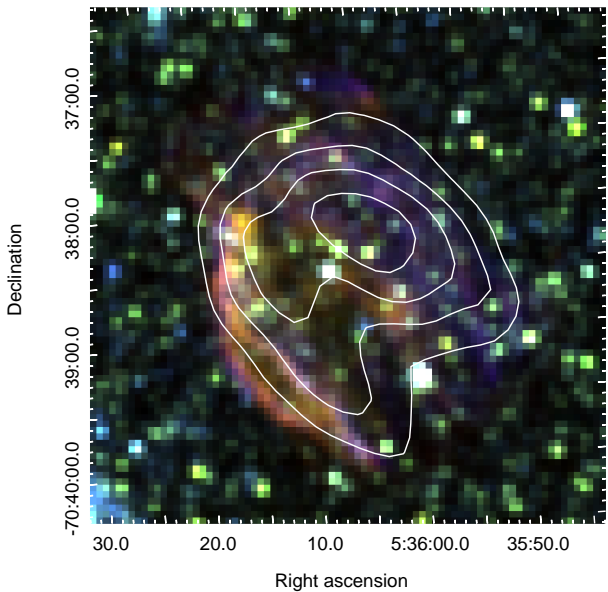}
\caption{MCELS composite optical image \textrm{(RGB =H$\alpha$,[S\textsc{ii}],[O\textsc{iii}])} of \SNR\ overlaid with 6~cm contours. The contours are 3, 6, 9 and 12$\sigma$ ($\sigma$ = 0.2~mJy/beam)}\label{mcels}
\end{center}
\end{figure}

We also used the Magellanic Cloud Emission Line Survey (MCELS) that was carried out with the 0.6~m University of Michigan/CTIO Curtis Schmidt telescope, equipped with a SITE $2048 \times 2048$\ CCD, which gave a field of 1.35\degr\ at a scale of 2.4\arcsec\,pixel$^{-1}$. Both the LMC and SMC were mapped in narrow bands corresponding to \Halpha, \OIII\ ($\lambda$=5007\,\AA), and \SII\ ($\lambda$=6716,\,6731\,\AA). All the data has been flux-calibrated and assembled into mosaic images, a small section of which is shown in Fig.~\ref{mcels}. Further details regarding the MCELS are given by \citet{pell2012} and at http://www.ctio.noao.edu/mcels. 

\section{Results and Discussion}

\SNR\ exhibits a horseshoe morphology at 6~cm (Fig.~\ref{extent}) centered at RA (J2000) = 5$^h$ 36$^m$ 07$^{s}$.7, Dec (J2000) = --70\degr38\arcmin20\arcsec.  We selected a one-dimensional intensity profile across the approximate major (NE--SW) and minor (SE--NW) axes (Fig.~\ref{extent}) at the 3$\sigma$ noise level (0.6 mJy) to estimate the spatial extent of the remnant. Its size at 6~cm is 130\arcsec$\times$130\arcsec\ with a 4\arcsec\ uncertainty in each direction (32 $\times$ 32 pc with a 1 pc uncertainty in each direction at the LMC distance of 50~kpc \citep{db08}). The remnant appears to be split into two limbs, the thicker, brighter limb toward the north-west and the thinner fainter limb towards the south-east.  

The optical emission from the MCELS correlates nicely with our 6~cm radio-continuum emission. The \SII\ and \Halpha\ emission are quite similar, with rather uniform emission spread over the remnant but exhibiting brightening on the south-east limb. \OIII\ emission is also found on this south-eastern limb (although thinner), however, unlike the \Halpha\ it is also found to show brightening towards the north-western limb. This is comparable to another LMC SNR that shares many similar properties (potential prompt Type~Ia, older, opposite \Halpha/\SII\ - \OIII), MCSNR J0508--6902 \citep{bozzetto2014}. The emission located towards the south-east of the remnant is a cluster of stars -- NGC 2056 and not likely associated with this SNR.

\begin{figure}[h]
\begin{center}
\includegraphics[scale=.5, trim=30 30 40 57,clip]{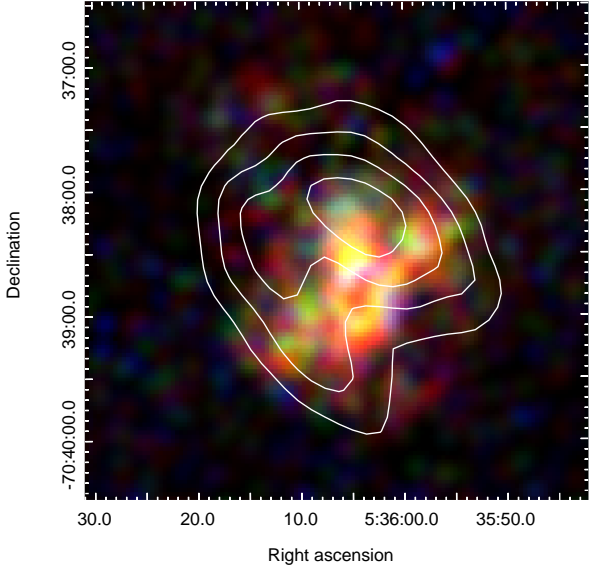}
\caption{Chandra three colour composite (red: 0.45--1.03 keV, green: 1.03--1.32 keV, blue: 1.32--2.2 keV) of \SNR\ overlaid with 6~cm radio contours of 3 6 9 \& 12$\sigma$ ($\sigma$ = 0.2~mJy)}\label{chandra}
\end{center}
\end{figure}

There has already been an extensive X-ray study done for this SNR. However, we have overlaid 6~cm radio contours on a three colour composite Chandra image (Obs. ID:~3908) to show the association between the X-ray and radio-continuum emission (Fig.~\ref{chandra}). We find that the radio-continuum emission loosely follows the low energy X-ray band (0.45--1.03~keV) and the  medium band X-ray emission is confined in the interior of the remnant. 

The spectral energy distribution (SED) of the remnant between 36~cm and 6~cm ($\alpha=-0.52\pm0.07$; based on the values in Table~\ref{tbl-flux}) shows the non-thermal nature of this remnant at radio-wavelengths. This value of $-0.5$ is typical spectral index of SNRs \citep{mathewson83,filipovic98}.

As we are unable to create a reliable 3~cm polarisation image (due to this wavelength being significantly affected by missing short spacing), we instead make use of the wide 2~GHz bandwidth from the 6~cm observations in order to carry out a comparative polarisation study. The 2048 channels at 6~cm were split into two even parts, each consisting of 1024 channels. Both images were convolved to the same resolution so that they could be compared.

Fractional polarisation ({\textit P}) was calculated for both 6~cm images using:\\
\vspace{2mm}

\begin{figure}[h!]
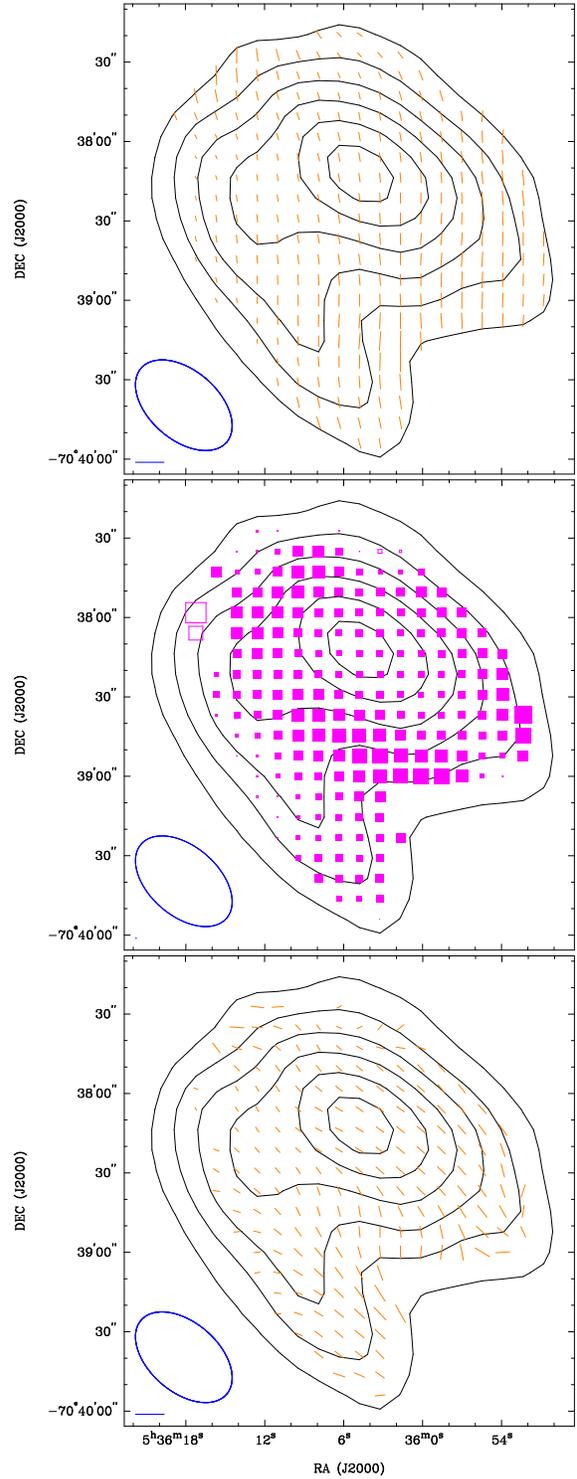

\begin{center}
\includegraphics[scale=.36, angle=-90, trim=0 0 50 0,clip]{0536-7038-6cm_electric}
\includegraphics[scale=.36, angle=-90, trim=0 0 50 0,clip]{0536-7038-6cm_rm}
\includegraphics[scale=.36, angle=-90, trim=0 0 0 0,clip]{0536-7038-0wavelength}
\caption{Top - electric field of \SNR\ at 6~cm. Middle - rotation measure, with filled boxes representing positive rotation and open boxes representing negative rotation. Bottom - Intrinsic magnetic field vectors with the polarised intensity based on 6~cm values. All images are overlaid with 6~cm contours of 3, 6, 9, 12, 15 \& 18$\sigma$ ($\sigma$ = 2 mJy/beam). The ellipse in the lower left corner represents the synthesised beamwidth of 42.7\arcsec$\times$26.1\arcsec and the line directly below (in the top and bottom images) shows a polarisation vector of 100\%
}\label{polar}
\end{center}
\end{figure}

{\textit P} = $\frac{\sqrt{S_{Q}^{2}+S_{U}^{2}}}{S_{I}}$\\

\noindent where $S_{Q}, S_{U}$ and $S_{I}$ are integrated intensities for \textit{Q}, \textit{U} and \textit{I} Stokes parameters. We estimate a mean fractional polarisation value of $P\cong$~35~$\pm$~8\% at 6~cm. This relatively high level of polarisation is (theoretically) expected for an SNR with a radio spectrum of less than $-0.5$ \citep{rolfs03}, which is in agreement with our spectral index of $\alpha \cong-0.52$. The structure of this polarisation can be seen in the top image of Fig.~\ref{polar}, where the electric field vectors at 6~cm have been superimposed on the 6~cm contours. These vectors are primarily facing the north-south direction, with slight change over the remnant, mostly found towards the norths and south limbs. We find a peak fractional polarisation for this SNR of $P=71$~$\pm$~25\%. This level of polarisation is comparatively higher than most LMC SNRs, and is on par with the peak value found for LMC SNR~J0455--6838 by \citet{crawford08}, of $\sim$70\%.

High levels of intrinsic fractional polarisation (with a maximum of $p=(3-3\alpha)/(5-3\alpha)$ = $\sim$70\% in a ordered field) from synchrotron emission is possible, and we do expect the magnetic field components perpendicular to compression to be amplified. However, such levels of polarisation are not expected to be observed due to the non-uniform regions of the polarised emission in addition to instrumental depolarisation and/or physical depolarisation effects outside the SNR. A possible explanation for these high values is given in \citet{dickel2000} where they pointed that the fractional polarisation can be artificially increased due to the polarised intensity having more fine scale structure than the total intensity, and therefore, the missing short spacing between antennas misrepresents the background level. 

Polarisation position angles were taken from both 6~cm polarisation images and used to estimate Faraday rotation for this SNR. The rotation measure for this remnant can be seen in Fig.~\ref{polar}, where positive rotation measure is shown by filled boxes and negative rotation by empty boxes. The rotation measure is predominately positive across the remnant, with a few negative values found towards the edge of the remnant. However, the polarised intensity where these negative values are situated is too low to accurately determine rotation measure, and are therefore, probably not real. The average rotation measure across the remnant is 237~rad~m$^{-2}$.

To see the intrinsic magnetic field of \SNR, we first de-rotate the 6~cm electric field vectors to their zero-wavelength position angle, and then rotate the vectors by 90 degrees to get the perpendicular magnetic field. The result of this can be seen in Fig.~\ref{polar}, where we find the magnetic field vectors overlaid on 6~cm contours. The vectors appear uniform across the field, with slight deviations along the east side and a couple on the northern most line. However, all lie within the 10--20\% of 6--9~$\sigma$ which makes them 1-2$\sigma$, and therefore not significant. The overall morphology suggests that \SNR\ went off in a region of uniformly NE--SW field and so compressed it more in the NW--SE direction.

From the position of \SNR\ at the surface brightness to diameter ($\Sigma$ - D) diagram ((D, $\Sigma$) = (32~pc, 3.6~$\times$~10$^{-21}$~W~m$^{-2}$~Hz$^{-1}$~sr$^{-1}$)) by \citet{berezhko04}, we can estimate that the remnant is likely to be an SNR in the late energy conserving phase, with an explosion energy between 0.25 and 1 $\times$ 10$^{51}$ ergs, which evolves in an environment of density $\sim$1 cm$^{-3}$.

\section{Conclusions}

We provide a radio-continuum study of \SNR, measuring an extent of 32 pc $\times$32 pc, strong polarisation across the remnant, with a mean of ~35\%, nice optical to radio association for the SNR. We also find the intrinsic magnetic field for this remnant, which uniformly follows the path of the compressed emission. We estimate a spectrum of $\alpha=-0.52$ between $\lambda$=6~cm \& 36~cm, typical of a mid to older remnant. The rotation measure across the remnant is predominately positive with a mean rotation measure of --237~rad~m$^{-2}$.

\acknowledgments

The Australia Telescope Compact Array is part of the Australia Telescope which is funded by the Commonwealth of Australia for operation as a National Facility managed by CSIRO. The MCELS is funded through the support of the Dean B. McLaughlin fund at the University of Michigan and through NSF grant 9540747. The scientific results reported in this article are based on observations made by the \emph{Chandra X-ray Observatory} (CXO). We thank the referee for numerous helpful comments that have greatly improved the quality of this paper.


%

%

\end{document}